\documentclass{IEEEtran}
\usepackage{amssymb}
\usepackage{amsfonts}
\usepackage{amsmath}
\usepackage{algorithm,algorithmic}
\usepackage{graphicx,subfigure,cite,multicol,multirow,diagbox,booktabs,array,stfloats}

\usepackage[dvips]{color}
\usepackage{float}
\ifCLASSINFOpdf
\else
\fi
\begin{document}

\title{Two High-Performance Amplitude Beamforming Schemes for Secure Precise Communication and Jamming with Phase Alignment
}
\author{Lingling Zhu,~Bofan Wu,~Lin Liu,~Jun Zou,\\~Jiayu Li,~Yuntian Wang,~Tong Shen,~Feng Shu
\thanks{Lingling Zhu,~Bofan Wu,~Lin Liu,~Jun Zou,~Jiayu Li,~Yuntian Wang,~Tong Shen,~Feng Shu are with the School of Electronic and Optical Engineering, Nanjing University of Science and Technology, 210094, CHINA (email: shufeng@njust.edu.cn).}
}
\maketitle

\begin{abstract}
To severely weaken the eavesdropper's ability to intercept confidential message (CM), precise jamming (PJ) is proposed by making use of the concept of secure precise wireless transmission (SPWT). Its basic idea is to focus  the transmit energy of artificial noise (AN) onto the neighborhood of eavesdropper (Eve) by using random subcarrier selection, directional modulation, phase alignment (PA), and amplitude beamforming (AB). By doing so, Eve will be seriously interfered with AN. Here, the conventional joint optimization of phase and amplitude is converted into two independent phase and amplitude  optimization problems. Considering PJ and SPWT require PA, the joint optimization problem reduces to an  amplitude  optimization problem. Then, two efficient AB schemes are proposed: leakage and maximizing receive power(Max-RP). Simulation results show that the proposed Max-RP and leakage AB methods perform much better than conventional equal AB (EAB) method in terms of both bit-error-rate and secrecy rate at medium and high signal-to-noise ratio regions. The performance difference between the two proposed leakage and Max-RP amplitude beamformers is trivial. Additionally, we also find the fact that all three AB schemes EA, Max-RP, and leakage can form two main peaks of AN and CM around Eve and the desired receiver, respectively.
\end{abstract}


\IEEEpeerreviewmaketitle
\section{Introduction}
In recent years, physical-layer security (PLS) has received significant attention in establishing secure wireless communications against interceptors\cite{wyner,wuyongpeng,majianping}. To achieve PLS and improve the secrecy rate (SR), beamforming, artificial noise (AN) and relay cooperation have been incorporated into secrecy network. As a PLS technology, directional modulation (DM) transmits the direction-dependent signals that are normal in the desired directions while disordered in all other directions\cite{yuanding,hujinsongDM,wuxiaomin}. Now, DM is widely studied and can be directly applied to millimeter wave and unmanned air vehicle (UAV) communication systems.

However, DM has a limitation in that the beam steering is fixed in an angle for all ranges. To overcome the angle dependence and distance independence characteristics of DM,  the authors in \cite{huAN,wuxiaomin1} proposed secure precise wireless transmission (SPWT) technology which focuses the transmit energy in a desired two-dimensional spatial section. SPWT can be considered as an effective combination of frequency diverse array (FDA) and DM technology. Specially, FDA \cite{FDA} exploited different frequency offsets to decouple the channels of Bob and Eve, which makes up for the above shortcoming of DM. In \cite{linjingran}, the authors optimized the frequency offsets  by block successive upper-bound minimization algorithm to ensure the maximum SR of proximal Bob and Eve scenario. To further consider the scenario where Eve is sensitive enough, the authors in \cite{qiubin} made full use of AN to achieve a higher SR. Moreover, the authors extended the frequency offset optimization investigation in the case of multiple eavesdroppers. Different from frequency offset optimization to maximize SR, the authors in \cite{shentong} studied the SPWT in the orthogonal frequency diversion multiplexing (OFDM) wireless communication system, where the subcarrier index was highly randomized by integer mod, ordering and block interleaving. With the assistance of singular value decomposition (SVD), a multi-beam DM scheme based on multi-carrier FDA was proposed, which reduces both implementing complexity and power consumption\cite{chengqian}.

Although beamforming schemes have been investigated intensively, almost all of the previous studies design the amplitude and phase simultaneously. In this letter, we propose a novel beamforming scheme for a communication system of PJ with artificial noise based on random-subcarrier-selection (RSS-PJ-AN), our main contributions in this paper are as follows:
\begin{enumerate}
 \item  To focus the AN energy on Eve, we extend the idea of SPWT to a precise jamming (PJ). By doing so, Eve will be degraded seriously. To achieve both SPWT and PJ, phase alignment (PA) is mandatory for the beamforming vectors of CM and AN. Then, a novel beamforming framework is proposed for SPWT and PJ. In other words, the beamforming design is decomposed into two parts: PA and amplitude beamforming (AB).

 \item  Using the proposed new beamforming framework, two AB schemes are proposed: leakage and maximum receive power (Max-RP).  Both AB of AN and CM are based on the two rules.  Simulation results show that, compared to equal AB (EAB), the proposed two AB schemes leakage and Max-RP perform better in the medium and high signal-to-noise ratio (SNR) regions in terms of  SR and bit-error-rate (BER) performance while they have the same performance as EAB in  the low SNR region.
     \end{enumerate}

The remainder of this paper is organized as follows. In Section \ref{Section 2}, we describe RSS-PJ-AN system model. Subsequently, we propose two AB methods: leakage-based and Max-RP in Section \ref{Section 3}. In Section \ref{Section 4}, the performance of the proposed scheme is numerically evaluated, and conclusions are given in Section \ref{Section 5}.

Notations: matrices, vectors, and scalars are denoted by letters of bold upper case, bold lower case, and lower case, respectively. Signs $(\cdot)^T$, $(\cdot)^H$ and $(\cdot)^{-1}$ denote matrix transpose, conjugate transpose and Moore-Penrose inverse, respectively. $\mathrm{diag}(a_1,a_2,\cdots,a_N)$ returns the diagonal concatenation. The operation $|\cdot|$ denotes modulus of a complex number. The symbol $\mathbf{I}_N$ denotes the $N \times N$ identity matrix.

\section{System Model}\label{Section 2}
\begin{figure}[h]
  \centering
  \includegraphics[width=0.3\textwidth]{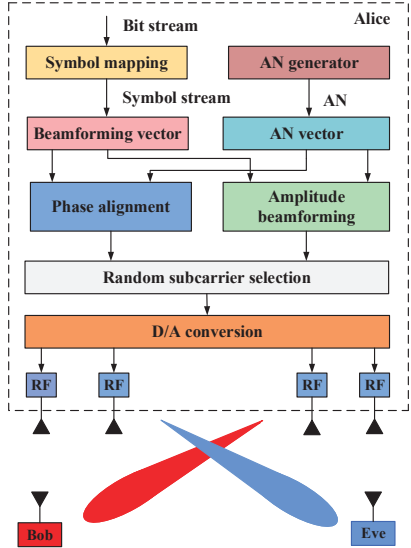}\\
  \caption{Block diagram for RSS-PJ-AN systems.}\label{Sys_Mod}
\end{figure}

Fig.~\ref{Sys_Mod} illustrates a typical architecture for RSS-PJ-AN system model consisting of an $N$-antenna uniform linear transmit array, a single-antenna  Bob and a single-antenna Eve. CM is transmitted towards Bob via randomly-selected multiple subcarriers from all-subcarrier set of OFDM. The all-subcarrier set of OFDM is $S_{sub} = \{f_m|f_m=f_c+ m\Delta f,  m=0,1,\ldots,N_S-1\}$, where $f_c$ is the carrier frequency and $\Delta f$ is the subchannel bandwidth. The subcarrier assigned to $n$-th antenna is $f_n$, where $f_n \in S_{sub}$\cite{wuxiaomin1}.

In this work, we assume that the channels between transmitter and receivers are LoP ones. The normalized steering vector for the transmit antenna array is given by
\begin{align}
\mathbf{h}(\theta,R) = \frac{1}{\sqrt{N}}[e^{j \Psi_0(\theta,R)},\cdots, e^{j \Psi_{N-1}(\theta,R)}]^T,
\end{align}
where $\Psi_n(\theta,R) = 2\pi f_n\frac{R-(n-1)d\cos\theta}{c} - 2\pi f_c\frac{R}{c}$ with $d$ being element spacing of uniform linear array (ULA) and $c$ being light speed. $\theta$ and $R$ are the direction angle and distance from Alice to receiver. Normally, frequency increment and carrier frequency can satisfy $N_S\Delta f  \ll f_c$. It is assumed that Bob and Eve are located at $(\theta_B,R_B)$ and $(\theta_E,R_E)$ with a high-resolution direction of arrival estimation \cite{Qinyaolu}, respectively.

The baseband transmit signal can be expressed as
\begin{align}
\mathbf{s} = \sqrt{\beta P_s}\mathbf{v}_{CM}x + \sqrt{(1-\beta)P_s}\mathbf{v}_{AN}z,
\end{align}
where $x$ is the CM  and $z$ is the AN both with average power constraint (i.e., $E[\left | x \right |^2]=1$, $E[\left | z \right |^2]=1$). $P_s$ is the transmit power of Alice and $\beta$ is the parameter that determines the power allocation between the CM and AN. $\mathbf{v}_{CM} $ and $\mathbf{v}_{AN}$ are the normalized CM beamforming vector and the normalized AN vector, respectively.

Considering that SPWT and PJ require $\mathbf{v}_{CM}$ to align with Bob ($\theta_B, R_B$) and $\mathbf{v}_{AN}$ to align with Eve ($\theta_E, R_E$), the n-th element of phase vector of $\mathbf{v}_{CM}$ and $\mathbf{v}_{AN}$ can be expressed as $\mathrm{ arg}(\mathbf{v}_{CM}(n)) = \Psi_n(\theta_B,R_B)$ and $\mathrm{arg}(\mathbf{v}_{AN}(n)) = \Psi_n(\theta_E,R_E)$, respectively. Here, we define $\mathbf{v}_{CM} = \mathbf{P}\mathbf{a}$ and $\mathbf{v}_{AN} = \mathbf{Q}\mathbf{b}$, where  $\mathbf{P}$ and $\mathbf{Q}$ are both diagonal matrix as $\mathbf{P}=\mathrm{ diag}[\mathrm{ arg}(\mathbf{v}_{CM}(0)),
\cdots, \mathrm{ arg }(\mathbf{v}_{CM}(N-1))] $ and $\mathbf{Q}=\mathrm{ diag}[\mathrm{ arg}(\mathbf{v}_{AN}(0)), \cdots, \mathrm{ arg} (\mathbf{v}_{AN}(N-1))]$, respectively. Meanwhile, $\mathbf{a}$ is the AB vector of $\mathbf{v}_{CM}$ and $\mathbf{b}$ is the AB vector $\mathbf{v}_{AN}$ (i.e., $|\mathbf{v}_{CM}(n)| = \mathbf{a}(n)$ and $|\mathbf{v}_{AN}(n)| = \mathbf{b}(n)$).
Accordingly, the received signal at Bob and Eve can be formulated as follows
\begin{align}
y(\theta_B,R_B)
&=\sqrt{g_d\beta P_s}\mathbf{h}^H(\theta_B,R_B)\mathbf{P}\mathbf{a}x \nonumber\\
&+\sqrt{g_d(1-\beta) P_s}\mathbf{h}^H(\theta_B,R_B)\mathbf{Q}\mathbf{b}z+n_B,
\end{align}
and
\begin{align}
y(\theta_E,R_E)
&=\sqrt{g_e\beta P_s}\mathbf{h}^H(\theta_E,R_E)\mathbf{P}\mathbf{a}x \nonumber\\
&+\sqrt{g_e(1-\beta) P_s}\mathbf{h}^H(\theta_E,R_E)\mathbf{Q}\mathbf{b}z+n_E,
\end{align}
where $n_B$ and $n_E$ are the additive white Gaussian noise (AWGN), distributed as $n\sim\mathcal{C}\mathcal{N}(0,\sigma^2)$ and $n\sim\mathcal{C}\mathcal{N}(0,\sigma^2)$, respectively. $g_d = R_0/R_B^2$ and $g_e = R_0/R_E^2$ denote  path loss coefficients from Alice to Bob and from Alice to Eve, respectively. $R_0$ is the reference distance which is set to 1m.

In what follows,  for the convenience of optimization,  $\mathbf{a}$ and $\mathbf{b}$ are relaxed as  complex optimization variables, and the corresponding element magnitudes of vectors $\mathbf{v}_{CM}$ and $\mathbf{v}_{AN}$ are set to be the element magnitudes  of optimal $\mathbf{a}$ and $\mathbf{b}$ in terms of some rules.
\section{Two Proposed AB Schemes}\label{Section 3}
In this section, to improve the SR performance,  two AB schemes leakage and Max-RP with EAB methods are proposed to optimize the amplitude parts of the beamforming vectors of AN and CM. Here, their phase parts are directly set to the negation of phases of channel steering vectors in order to form the main peaks of CM and AN at Bob and Eve, respectively.
\subsection{Proposed Leakage-Based AB Method}\label{Section 3A}
Making use of leakage criterion, we use the maximizing signal-to-leakage-noise ratio (Max-SLNR) method to optimize the vector $\mathbf{a}$, which can be obtained as
\begin{align}\label{P1}
&\max_{\mathbf{a}}~~~~\rm SLNR(\mathbf{a})\\
&~\rm s.t. ~~~~~~\mathbf{a}^H\mathbf{a}=1,\nonumber
\end{align}
where
\begin{align}
\rm SLNR(\mathbf{a}) = \frac{g_d\mathbf{a}^H \mathbf{P}^H\mathbf{h}(\theta_B, R_B)\mathbf{h}^H(\theta_B, R_B)\mathbf{P}\mathbf{a} }{\mathbf{a}^H(g_e \mathbf{P}^H\mathbf{h}(\theta_E, R_E)\mathbf{h}^H(\theta_E, R_E)\mathbf{P} +\frac{\sigma^2}{\beta P_s}\mathbf{I}_N)\mathbf{a}}.
\end{align}
which yields  the vector $\mathbf{a}$ being the eigenvector corresponding to the largest eigenvalue of matrix
\begin{align}
&\left[ g_e\mathbf{P}^H\mathbf{h}(\theta_E, R_E)\mathbf{h}^H(\theta_E, R_E)\mathbf{P} +\frac{\sigma^2}{\beta P_s}\mathbf{I}_N\right]^{-1} \nonumber\\
&\times g_d\mathbf{P}^H\mathbf{h}(\theta_B, R_B)\mathbf{h}^H(\theta_B, R_B)\mathbf{P}.
\end{align}
Since $\mathbf{a}$ is a complex vector, the magnitude $|\mathbf{v}_{CM}(n)|$ of element $n$ of vector $\mathbf{v}_{CM}$ is chosen to be $|\mathbf{a}(n)|$. Next, the AN is viewed as the useful signal of Eve. Let  $\mathbf{b}$ be the optimization vector, similarly, we have
\begin{align}\label{P2}
&\max_{\mathbf{b}}~~~~\rm SLNR(\mathbf{b})\\
&~\rm s.t. ~~~~~~\mathbf{b}^H\mathbf{b}=1,\nonumber
\end{align}
where
\begin{align}\label{ANSNR}
\rm SLNR(\mathbf{b}) = \frac{g_e\mathbf{b}^H \mathbf{Q}^H\mathbf{h}(\theta_B, R_B)\mathbf{h}^H(\theta_B, R_B)\mathbf{Q}\mathbf{b} }{\mathbf{b}^H(g_d \mathbf{Q}^H\mathbf{h}(\theta_E, R_E)\mathbf{h}^H(\theta_E, R_E)\mathbf{Q} +\frac{\sigma^2}{\beta P_s}\mathbf{I}_N)\mathbf{b}}.
\end{align}
which results in the optimal value of $\mathbf{b}$ being the generalized eigenvector corresponding to the largest normalized eigenvalue of matrix
\begin{align}
&\left[ g_d\mathbf{Q}^H\mathbf{h}(\theta_B, R_B)\mathbf{h}^H(\theta_B, R_B)\mathbf{Q} +\frac{\sigma^2}{(1-\beta) P_s}\mathbf{I}_N\right]^{-1}\nonumber\\
&\times g_e\mathbf{Q}^H\mathbf{h}(\theta_E, R_E)\mathbf{h}^H(\theta_E, R_E)\mathbf{Q}.
\end{align}
Likewise, the AB vector of $\mathbf{v}_{AN}$ is taken to be $[|\mathbf{b}(0)|,|\mathbf{b}(1)|, \cdots, |\mathbf{b}(N-1)|]^T$.
\subsection{Proposed Max-RP-based AB Method}\label{Section 3B}
Now, we  turn to a new rule. By maximizing the receive power of CM at Bob, the AB part of $\mathbf{v}_{CM}$  are constructed.  The corresponding optimization problem of Max-RP is given by
\begin{align}\label{P3}
&\max_{\mathbf{a}}~~~~~~~~\mathbf{a}^H\mathbf{P}^H\mathbf{h}(\theta_B,R_B)\mathbf{h}^H(\theta_B,R_B)\mathbf{P}\mathbf{a}\\
&~\rm s.t. ~~~~~~~~~~  \mathbf{h}^H(\theta_E,R_E)\mathbf{P}\mathbf{a}=0,\nonumber
\end{align}
where  the constraint $\mathbf{h}^H(\theta_E,R_E)\mathbf{P}\mathbf{a}=0$ forces CM to transmit on the null space of $\mathbf{h}^H(\theta_E,R_E)\mathbf{P}$. To address the above problem, the singular-value decomposition (SVD) of $\mathbf{h}^H(\theta_E,R_E)\mathbf{P}$ is computed as $\mathbf{h}^H(\theta_E,R_E)\mathbf{P} = [U_e] (\mathbf{\sum} ^{(1)}_e ~~ \mathbf{0})[\mathbf{V}^{(1)}_e ~~\mathbf{V}^{(0)}_e]^H$, where $\sum^{(1)}_e$ is a complex number, and $\mathbf{V}^{(0)}_e$ consists of the last $(N - 1)$ right singular vectors corresponding to $N-1$ zero singular values. Define $\mathbf{F}_e = \mathbf{V}^{(0)}_e$, and $\mathbf{a}= \mathbf{F}_e\mathbf{u}$, then the optimization problem in (\ref{P3}) is converted into
\begin{align}\label{P4}
&\max_{\mathbf{u}}~~~~~~~\mathbf{u}^H\mathbf{F}_e^H\mathbf{P}^H\mathbf{h}(\theta_B,R_B)\mathbf{h}^H(\theta_B,R_B)\mathbf{P}\mathbf{F}_e\mathbf{u}\\
&~\rm s.t. ~~~~~~~~~ \mathbf{u}^H\mathbf{u}=1,\nonumber
\end{align}
which means that $\mathbf{u}$ is the eigenvector corresponding to the largest eigenvalue of matrix $\mathbf{F}_e^H\mathbf{P}^H
\mathbf{h}(\theta_B,R_B)\mathbf{h}^H(\theta_B,R_B)\mathbf{P}\mathbf{F}_e$.
 The design of $\mathbf{a}$ has been finished. Then, the AB vector of $\mathbf{v}_{CM}$ is $[|\mathbf{a}(0)|, \cdots, |\mathbf{a}(N-1)|]^T$.

Similar to (\ref{P3}), we can readily obtain
\begin{align}\label{P5}
&\max_{\mathbf{b}}~~~~~~~~\mathbf{b}^H\mathbf{Q}^H\mathbf{h}(\theta_E,R_E)\mathbf{h}^H(\theta_E,R_E)\mathbf{Q}\mathbf{b}\\
&~\rm s.t. ~~~~~~~~~  \mathbf{h}^H(\theta_B,R_B)\mathbf{Q}\mathbf{b}=0.\nonumber
\end{align}
The SVD of $\mathbf{h}^H(\theta_B,R_B)\mathbf{Q}$ is $\mathbf{h}^H(\theta_B,R_B)\mathbf{Q} = [U_b] (\mathbf{\sum} ^{(1)}_b ~~ \mathbf{0})[\mathbf{V}^{(1)}_b ~~\mathbf{V}^{(0)}_b]^H$. Define $\mathbf{F}_b = \mathbf{V}^{(0)}_b$, and $\mathbf{b} = \mathbf{F}_b\mathbf{w}$, then the optimization problem in (\ref{P5}) can also be expressed as
\begin{align}\label{P6}
&\max_{\mathbf{w}}~~~~~~~\mathbf{w}^H\mathbf{F}_b^H\mathbf{Q}^H\mathbf{h}(\theta_E,R_E)\mathbf{h}^H(\theta_E,R_E)\mathbf{Q}\mathbf{F}_b\mathbf{w}\\
&~\rm s.t. ~~~~~~~~~ \mathbf{w}^H\mathbf{w}=1,\nonumber
\end{align}
where $\mathbf{w}$ is the eigenvector corresponding to the largest eigenvalue of matrix $\mathbf{F}_b^H\mathbf{Q}^H\mathbf{h}(\theta_E,R_E)
\mathbf{h}^H(\theta_E,R_E)\mathbf{Q}\mathbf{F}_b$. Then the design of $\mathbf{b}$ and $|\mathbf{v}_{AN}(n)| = |\mathbf{b}(n)|$ have been completed.
\section{Simulations and Discussions}\label{Section 4}
In our simulation, system parameters are set as follows: quadrature phase shift keying (QPSK) modulation, the total signal bandwidth is 5MHz, $f_c$ = 3GHz, the number of total subcarriers $N_S = 1024$, $N = 32$, $d= \lambda /2$, $\beta =0.5$, $\sigma^2= -60dBm$, $(\theta_B, R_B)=(30^{\circ}, 650 \rm{m})$ and $(\theta_E,R_E)=(100^{\circ}, 550 \rm{m})$.
\begin{figure*}[htbp]
\centering
\subfigure[EAB.]{
\begin{minipage}[t]{0.31\linewidth}
\centering
\includegraphics[width=5.3cm]{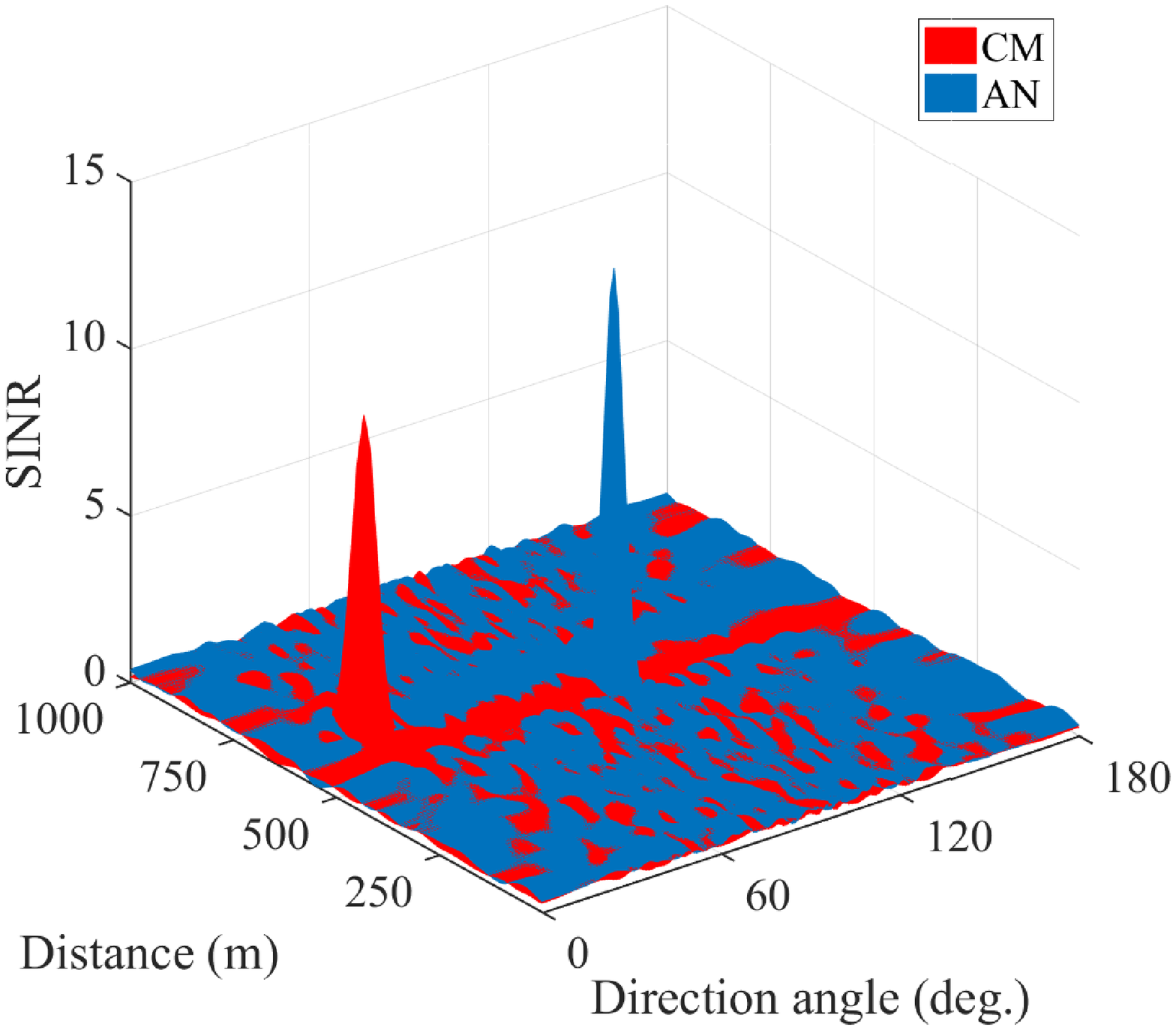}
\end{minipage}%
}%
\subfigure[Proposed leakage-based AB.]{
\begin{minipage}[t]{0.31\linewidth}
\centering
\includegraphics[width=5.3cm]{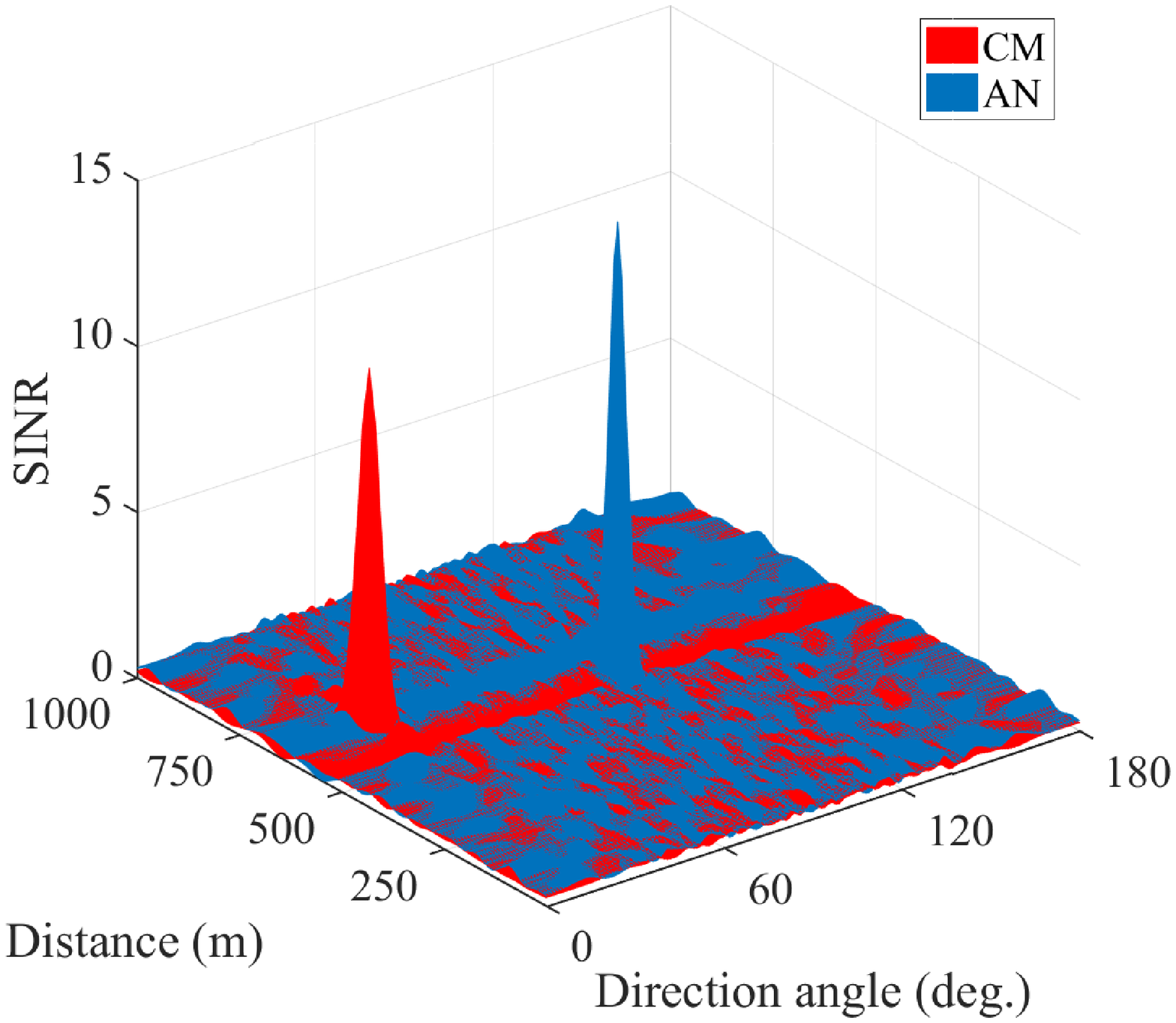}
\end{minipage}%
}%
\subfigure[Proposed Max-RP-based AB.]{
\begin{minipage}[t]{0.31\linewidth}
\centering
\includegraphics[width=5.3cm]{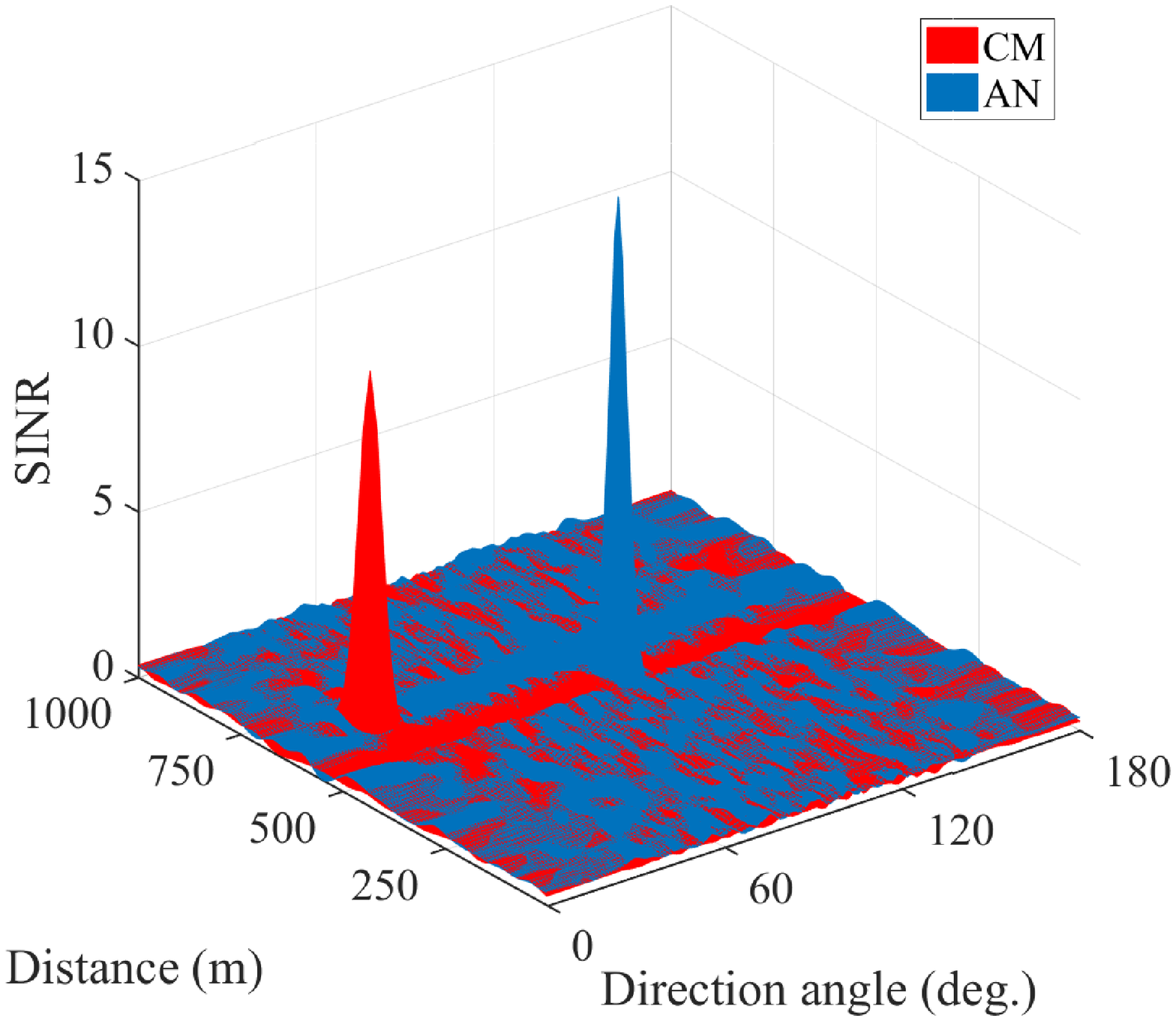}
\end{minipage}%
}%
\centering
\caption{ 3D  surfaces of SINR versus direction angle and distance of EAB, proposed leakage-based AB and proposed Max-RP-based AB.
}\label{SINR-ANSNR}
\end{figure*}

Fig.~\ref{SINR-ANSNR} plots the three-dimensional (3D) performance surface of signal-to-interference-plus-noise ratio (SINR) versus direction angle $\theta$ and distance $R$ of proposed methods for SNR = 14dB, where the conventional EAB method is used as a performance reference. Here, CM is a useful signal for Bob, and AN is a useful signal for Eve. Observing three parts in Fig.~\ref{SINR-ANSNR}, two peaks of CM and AN are only around Bob $ (30^{\circ}, 650 \rm{m})$ and Eve $(100^{\circ}, 550 \rm{m})$, respectively. Additionally, we also find a fact that the main peaks of Bob and Eve of the proposed methods are much higher than those of EAB, which means that the proposed methods have a better SINR performance.

\begin{figure}[h]
  \centering
  \includegraphics[width=0.4\textwidth]{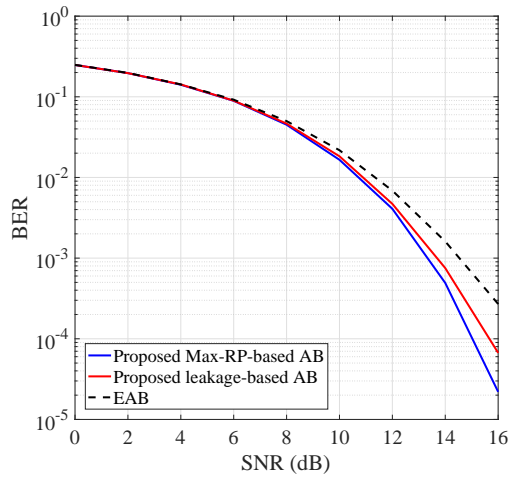}\\
  \caption{Curves of BER with EAB, proposed leakage-based AB and proposed Max-RP-based AB versus SNR.
  }\label{BER}
\end{figure}

Fig.~\ref{BER} shows the curves of BER versus the SNR, for the two proposed AB schemes described in Section-III. The figure illustrates that our proposed methods outperform the EAB. Particularly in the high SNR region, the BER performance of our proposed schemes is about one order of magnitude, better than that of the EAB method.

\begin{figure}[h]
  \centering
  \includegraphics[width=0.4\textwidth]{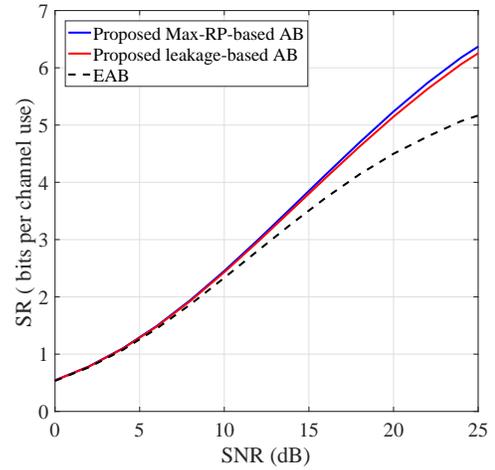}\\
  \caption{Curves of SR with EAB, proposed leakage-based AB and proposed Max-RP-based AB versus SNR.
}\label{SR}
\end{figure}

In what follows, we also evaluate the performance of the proposed AB schemes from SR aspect. Fig.~\ref{SR} demonstrates the curves of SR versus SNR of the proposed AB methods. From Fig.~\ref{SR}, it is noted that the SR performance of the three method are close to each other in the low SNR region. As SNR grows, the performance gap between the proposed schemes and EAB method becomes larger, which greatly improves the security of communication system.
\section{Conclusion}\label{Section 5}
In this paper, we mainly proposed two high-performance AB schemes: leakage and Max-RP to enhance the PLS of wireless communication. Simulation results showed that our proposed two AB schemes behaved better than EAB method in terms of BER and SR.
\ifCLASSOPTIONcaptionsoff
  \newpage
\fi
\bibliographystyle{IEEEtran}
\bibliography{cite}
\end{document}